\documentclass{elsarticle}

\usepackage{graphicx}
\usepackage{bm}
\usepackage{amssymb}
\usepackage{amsmath}
\usepackage{amsfonts}

\begin{document}

\begin{frontmatter}

\title{Acoustic-gravity waves in quasi-isothermal atmospheres with a random vertical temperature profile}


\author[mymainaddress,mysecondaryaddress]{V. M. Lashkin \corref{mycorrespondingauthor}}
\cortext[mycorrespondingauthor]{Corresponding author}
\ead{vlashkin62@gmail.com}

\author[mysecondaryaddress]{O. K. Cheremnykh}

\address[mymainaddress]{Institute for Nuclear Research, Pr. Nauki 47, Kyiv 03028,
Ukraine}
\address[mysecondaryaddress]{Space Research Institute, Pr. Glushkova 40
k.4/1, Kyiv 03187,  Ukraine}





\begin{abstract}
We study acoustic-gravity waves in a quasi-isothermal atmosphere
in the presence of a weak random addition to the vertical
temperature profile, which simulates the real atmosphere of the
Earth at altitudes greater than $\sim 200$ km. The resulting
stochastic equation is closed in the Bourret approximation. The
poles of the obtained mean Green's function determine the
generalized dispersion relation for acoustic-gravity waves. Two
particular cases are considered: random inhomogeneities in the
form of white noise ($\delta$-correlated in space) and the
opposite case of a $\delta$-shaped noise spectrum. In both cases,
instability of acoustic-gravity waves is predicted and the
corresponding instability growth rates are determined.
\end{abstract}

\end{frontmatter}

\section{Introduction}

Acoustic-gravity waves (AGWs) in the atmosphere of the Earth, in
the solar atmosphere, as well as in planetary atmospheres have
been the subject of a large number of experimental and theoretical
studies for many years
\cite{Eckart1960,Hines1960,Tolstoy1967,Liu1974,Beer1974,Gill1982}.
The sources of AGWs in the Earth's atmosphere and ionosphere can
be earthquakes, volcanic activity, tornadoes, thunderstorms, solar
eclipses, precipitation of charged particles and dissipation of
currents in the polar regions, magnetospheric storms, substorms,
strong ground explosions, and rocket launches. The results of
numerous experimental studies have shown that AGWs make a
significant contribution to the dynamics and energy of the Earth's
atmosphere, providing an effective interaction between different
altitude levels. These waves play a significant role in the
formation of atmospheric convection and turbulence. The study of
AGWs is also motivated by the need to obtain accurate predictions
of atmospheric dynamics under various meteorological conditions.

Despite the fact that the main properties of AGWs have been
studied and described in a number of books
\cite{Gossard1975,Sutherland2015} and reviews
\cite{Tolstoy1963,Francis1975}, and some progress has been made
both in the linear \cite{Collins1999,Walter2003,Cheremnykh2021}
and nonlinear \cite{Kaladze2008,Stenflo2009,Izvekova2015} theory
of these waves,  AGWs continue to attract a considerable
theoretical interest \cite{Misra2019,Gavrilov2021,Cheremnykh2022}.
Most of the results on  AGW theory were obtained in the model of
an isothermal atmosphere, in which the temperature is considered
to be a constant value that does not change with altitude. In the
real atmosphere of the Earth, the temperature profile is a very
complex and time-varying function of the vertical coordinate $z$
and, generally speaking, cannot be reduced to any deterministic
form. In addition, the temperature profile significantly depends
on latitude, daytime or nighttime, solar activity etc.
Nevertheless, at altitudes exceeding $\sim 200$ km, the atmosphere
can be considered quasi-isothermal \cite{Gavrilov2018,Picone2002}
and the model of an isothermal atmosphere can be used to describe
wave disturbances \cite{Gossard1975,Francis1975,Cheremnykh2021}.
In this paper, we focus on studying the propagation of
acoustic-gravity waves in a quasi-isothermal atmosphere, taking
into account a small random addition to the vertical temperature
profile. Such an addition can be caused, in particular, by
turbulent (sporadic) temperature fluctuations. We obtain a
stochastic equation for small-amplitude perturbations in a
quasi-isothermal atmosphere, which consistently takes into account
weak fluctuations in the vertical temperature profile. Using
statistical closure in the Bourret approximation
\cite{Burre1962,Kravtsov1989} that corresponds to a small random
addition, we obtain a dispersion relation describing the AGWs in
such a quasi-isothermal atmosphere. In the general case the
dispersion relation depends on the form of the autocorrelation
function of a random term. The dispersion equation is simplified
and amenable to analytical analysis in two particular cases of the
form of the autocorrelation function of a random process, white
noise and a narrow spectrum of noise concentrated near some
characteristic wave number.

\section{ Model Equations}

We begin with the linearized equations of motion for a
compressible non-rotating atmosphere \cite{Pedlosky1986}
\begin{equation}
\label{eq1} \frac{\partial \mathbf{u}^{'}}{\partial
t}+\nabla_{\perp}\varphi^{'}=0,
\end{equation}

\begin{equation}
\label{eq2} \frac{\partial w^{'}}{\partial t}+\left(\frac{\partial
\varphi^{'}}{\partial
z}-\frac{N^{2}}{g}\right)\varphi^{'}-g\frac{\theta^{'}}{\bar{\theta}}=0,
\end{equation}

\begin{equation}
\label{eq3} \frac{1}{c_{s}^{2}}\frac{\partial
\varphi^{'}}{\partial t}+\left(\frac{\partial
\varphi^{'}}{\partial z}-\frac{1}{\gamma
H}\right)w^{'}+\nabla_{\perp}\cdot\mathbf{u}^{'}=0,
\end{equation}

\begin{equation}
\label{eq4} \frac{\partial}{\partial
t}\frac{\theta^{'}}{\bar{\theta}}+\frac{N^{2}}{g}w^{'}=0.
\end{equation}
The first two equations are the horizontal and vertical (along
$z$-axis) momentum equations, respectively, the third is the
continuity equation and the last is the heat equation. Here
$\mathbf{u}$ and $w$ are the horizontal and vertical velocities,
and the quantities $\varphi$, $\theta$ (potential temperature),
$N$ (Brunt-V\"{a}is\"{a}l\"{a} frequency), $H$ (scale height) and
$c_{s}$ (sound speed) are defined as follows
\begin{gather}
\varphi=p/\bar{\rho}, \quad \theta=T (\bar{p}/p)^{\kappa} , \quad
N^{2}=d \ln \bar{\theta}/dz, \nonumber
\\
H=RT/g, \quad c_{s}^{2}=\gamma RT, \label{eq5}
\end{gather}
where $p$ and $\rho$ are the pressure and density respectively,
$\kappa = R/c_{p}$, $\gamma = c_{p}/c_{v}$, $R$ is the gas
constant, $g$ is the gravitational acceleration, $c_{p}$ and
$c_{v}$ are, respectively, the specific heats at constant pressure
and volume. Overbars refer to basic-state quantities and primes to
disturbance quantities. Supposing that the medium is stratified
only in altitude $z$, and is uniform horizontally, we introduce
the corresponding Fourier transforms as
\begin{equation}
\label{fourier}
(\mathbf{u}^{'},w^{'},\varphi^{'},\theta^{'})(\mathbf{r}_{\perp},z,t)=\int
(\mathbf{u},w,\varphi,\theta)\mathrm{e}^{i\mathbf{k}_{\perp}\cdot
\mathbf{r}_{\perp}-i\omega t}d^{2}\mathbf{k}_{\perp}\,d\omega,
\end{equation}
where $\omega$ and $\mathbf{k}_{\perp}$ are the frequency and
horizontal wave vector respectively. Eliminating $\mathbf{u}$ from
the Fourier transforms of equations (\ref{eq1}) and (\ref{eq3})
gives
\begin{equation}
\label{eq6}
\left(k_{\perp}^{2}-\frac{\omega^{2}}{c_{s}^{2}}\right)\varphi-i\omega\left(\frac{d}{d
z}-\frac{1}{\gamma\bar{H}}\right)w=0,
\end{equation}
and likewise eliminating $\theta$ from equations (\ref{eq2}) and
(\ref{eq4}) gives
\begin{equation}
\label{eq7} \left(N^{2}-\omega^{2}\right)w-i\omega\left(\frac{d}{d
z}-\frac{N^{2}}{g}\right)\varphi=0.
\end{equation}
Finally, eliminating $\varphi$ from (\ref{eq6}) and (\ref{eq7}),
we get the equation
\begin{equation}
\omega^{2}\left(\frac{d}{d
z}-\frac{N^{2}}{g}\right)\left(\frac{c_{s}^{2}}{k_{\perp}^{2}c_{s}^{2}-\omega^{2}}\right)
\left(\frac{d}{d z}-\frac{1}{\gamma
\bar{H}}\right)w+\left(N^{2}-\omega^{2}\right)w=0,
\end{equation}
which, taking into account (\ref{eq5}), can be rewritten as
\begin{gather}
\frac{d^{2}w}{dz^{2}}-\left[\frac{1}{H}
+\frac{k_{\perp}^{2}(c_{s}^{2})\prime}{(k_{\perp}^{2}c_{s}^{2}-\omega^{2})}\right]\frac{dw}{dz}
\nonumber \\
+\left[\left(\frac{\omega_{b}^{2}}{\omega^{2}}-1\right)k_{\perp}^{2}+\frac{\omega^{2}}{c_{s}^{2}}
+\frac{gk_{\perp}^{4}(c_{s}^{2})\prime}{\omega^{2}(k_{\perp}^{2}c_{s}^{2}-\omega^{2})}\right]w=0,
\end{gather}
where $\omega_{b}^{2}=(\gamma-1)g^{2}/c_{s}^{2}$ is the square of
Brunt-V\"{a}is\"{a}l\"{a} frequency corresponding to isothermal
atmosphere, and we have introduced the notation
$(c_{s}^{2})\prime=d c_{s}^{2}/dz$. Further, we restrict ourselves
to the case of not too large temperature gradients and consider
the case when
\begin{equation}
\label{condition}
\frac{k_{\perp}^{2}(c_{s}^{2})\prime}{(k_{\perp}^{2}c_{s}^{2}-\omega^{2})}\ll
\frac{1}{H}=\frac{\gamma g}{c_{s}^{2}}.
\end{equation}
This condition implies, in particular, that we consider the
frequency range far enough from the resonance
$\omega^{2}=k_{\perp}^{2}c_{s}^{2}$ corresponding to the Lamb
surface wave. Then, introducing $V(z)=w(z)\exp (-z/2H)$ we have
\begin{equation}
\label{main} \frac{d^{2}V}{dz^{2}}
+\left[\left(\frac{\omega_{b}^{2}}{\omega^{2}}-1\right)k_{\perp}^{2}+\frac{\omega^{2}}{c_{s}^{2}}
-\frac{\omega_{a}^{2}}{c_{s}^{2}}
+\frac{k_{\perp}^{4}c_{s}^{2}\omega_{b}^{2}}{\omega^{2}(k_{\perp}^{2}c_{s}^{2}-\omega^{2})}\varepsilon
(z)\right]V=0,
\end{equation}
where $\omega_{a}=c_{s}/2H$ is the so-called acoustic cutoff
frequency, and we have introduced the dimensionless function
$\varepsilon (z)=g(c_{s}^{2})\prime/(c_{s}^{2}\omega_{b}^{2})$. In
an isothermal atmosphere $\varepsilon (z)=0$, and then taking
$V\sim\exp (ik_{z}z-i\omega t)$ in (\ref{main}), we obtain the
known dispersion relation for acoustic-gravity waves
\cite{Eckart1960,Hines1960}
\begin{equation}
\label{disp-free}
F(\omega,\mathbf{k}_{\perp},k_{z})\equiv\omega^{4}-\omega^{2}(\omega_{a}^{2}+k^{2}c_{s}^{2})
+\omega^{2}_{b}k_{\perp}^{2}c_{s}^{2}=0,
\end{equation}
where $k^{2}=k_{\perp}^{2}+k_{z}^{2}$, and $\mathbf{k}_{\perp}$
and $k_{z}$ are the components of $\mathbf{k}$ across and along
the vertical direction. Equation (\ref{disp-free}) gives two
branches of oscillations,
\begin{equation}
\label{AGW-plus-minus}
\omega_{\pm}^{2}=\frac{\omega_{a}^{2}+k^{2}c_{s}^{2}}{2}\pm
\left[(\omega_{a}^{2}+k^{2}c_{s}^{2})^{2}-4\omega_{b}^{2}k_{\perp}^{2}c_{s}^{2}\right]^{1/2},
\end{equation}
with the high-frequency branch $\omega_{+}>\omega_{a}$,  and the
low-frequency branch $\omega_{-}<\omega_{b}$. Correspondingly,
waves satisfying these two distinct branches are usually called
acoustic waves and internal gravity waves respectively. The case
$\varepsilon (z)\neq 0$ corresponds to non-isothermal atmosphere.

Note that in this work we restrict ourselves to the case when we
can neglect the rotation of the Earth, that is, the Coriolis
force. This is certainly justified for the acoustic branch, since
$\omega_{a}/\Omega_{0}\sim 10^{2}$, where $\Omega_{0}$ is the
Earth's angular velocity. For gravity waves, we exclude from
consideration the case of very small wave numbers
$k\ll\Omega_{0}/c_{s}$, so that $\omega_{-}\gg \Omega_{0}$. In
addition, we also neglect the Ampere force, since at the
considered altitudes (E-layer) the effect of the geomagnetic field
is of the same order as the effect due to the Coriolis force
\cite{Kaladze2008}. The influence of the Coriolis force on the
propagation of AGWs was considered in detail in
\cite{Misra_Coriolis2021}. The effect of the Ampere force on
gravity waves in a weakly ionized Earth's ionosphere was studied
in \cite{Misra_Amper1-2022,Misra_Amper2022}.

We consider $\varepsilon (z)$ as a zero-mean
($\langle\varepsilon(z)\rangle=0$) homogeneous Gaussian random
field with the autocorrelation function
\begin{equation}
\label{correl}
\langle\varepsilon(z)\varepsilon(z^{'})\rangle=B(z-z^{'}),
\end{equation}
where the angular brackets $\langle \dots \rangle$ means ensemble
averaging. Thus, $\varepsilon(z)$ has the meaning of a random
addition to the vertical temperature profile. In the following, we
will consider the case corresponding to a quasi-isothermal
atmosphere, i.e. assume $B\ll 1$.

\section{Dispersion relation in the Bourret approximation}

Due to  explicit dependence on $z$, equation (\ref{main}), using
the convolution relation, can be written in the Fourier space in
the form
\begin{gather}
F(\omega,\mathbf{k}_{\perp},k_{z})V(\omega,\mathbf{k}_{\perp},k_{z})
+\frac{\omega^{2}_{b}k_{\perp}^{4}c_{s}^{4}}{(k_{\perp}^{2}c_{s}^{2}-\omega^{2})} \nonumber \\
\times\int
\varepsilon(k_{z,1})V(\omega,\mathbf{k}_{\perp},k_{z,2})\delta
(k-k_{z,1}-k_{z,2})\,dk_{z,1}dk_{z,2}=0, \label{general1}
\end{gather}
where $\varepsilon(k_{z})$ and
$V(\omega,\mathbf{k}_{\perp},k_{z})$ are the Fourier transforms of
$\varepsilon(z)$ and $V(\mathbf{r},t)$ respectively, and
$F(\omega,\mathbf{k}_{\perp},k_{z})$ is determined by
(\ref{disp-free}). Our goal is to obtain an equation for the mean
Green function $\langle G \rangle=\delta \langle V\rangle /\delta
\eta$, where $\eta$ is  source in the right-hand side of equation
(\ref{general1}). The poles of the Green function are known to
determine the spectrum of collective excitations, that is, the
dispersion relation. In order to avoid cumbersome expressions, we
introduce in some places (where this does not lead to a
misunderstanding) the notation
$q=(\omega,\mathbf{k}_{\perp},k_{z})$. Introducing the source
$\eta (\omega,\mathbf{k}_{\perp},k_{z})$ in the right-hand side of
(\ref{general1}) and taking the functional derivative
$\delta/\delta\eta (\omega^{'},\mathbf{k}_{\perp}^{'},k_{z}^{'})$,
we then have
\begin{gather}
F(q)G(q,q^{'})
+\frac{\omega^{2}_{b}k_{\perp}^{4}c_{s}^{4}}{(k_{\perp}^{2}c_{s}^{2}-\omega^{2})} \nonumber \\
\times\int
\varepsilon(k_{z,1})G(\omega,\mathbf{k}_{\perp},k_{z,2},q^{'})\delta
(k-k_{z,1}-k_{z,2})\,dk_{z,1}dk_{z,2}=\delta (q-q^{'}),
\label{general2}
\end{gather}
where $G(q,q^{'})=\delta n(q)/\delta \eta (q^{'})$, by definition,
is the Green function of equation (\ref{general2}), and we have
taken into account that $\delta \eta(q)/\delta \eta (q^{'})=\delta
(q-q^{'})$. Representing the Green function as a sum of the
average and fluctuating parts
\begin{equation}
\label{sum} G(q,q^{'})=\langle
G(q,q^{'})\rangle+\tilde{G}(q,q^{'}),
\end{equation}
substituting it into equation (\ref{general2}) and averaging, one
obtains
\begin{gather}
F(q)\langle G(q,q^{'})\rangle
+\frac{\omega^{2}_{b}k_{\perp}^{4}c_{s}^{4}}{(k_{\perp}^{2}c_{s}^{2}-\omega^{2})} \nonumber \\
\times\int
\langle\varepsilon(k_{z,1})\tilde{G}(\omega,\mathbf{k}_{\perp},k_{z,2},q^{'})\rangle\delta
(k-k_{z,1}-k_{z,2})\,dk_{z,1}dk_{z,2}=\delta (q-q^{'}).
\label{general3}
\end{gather}
Subtracting equation (\ref{general3}) from equation
(\ref{general2}), we get
\begin{gather}
F(q) \tilde{G}(q,q^{'})
+\frac{\omega^{2}_{b}k_{\perp}^{4}c_{s}^{4}}{(k_{\perp}^{2}c_{s}^{2}-\omega^{2})}
\int \varepsilon(k_{z,1})\langle
G(\omega,\mathbf{k}_{\perp},k_{z,2},q^{'})\rangle \nonumber
\\
\times\delta (k-k_{z,1}-k_{z,2})\,dk_{z,1}dk_{z,2} + \int [
\varepsilon(k_{z,1})
\tilde{G}(\omega,\mathbf{k}_{\perp},k_{z,2},q^{'})
\nonumber \\
-\langle \varepsilon(k_{z,1})
\tilde{G}(\omega,\mathbf{k}_{\perp},k_{z,2},q^{'})\rangle]\delta
(k-k_{z,1}-k_{z,2})\,dk_{z,1}dk_{z,2}=0.
 \label{general4}
\end{gather}
In the Bourret approximation
\cite{Burre1962,Kravtsov1989,Baydoun2015,Grinevich1997}, which is
justified when the intensity of the noise is small enough so that
$B(z-z^{'})\ll 1$, we can neglect the term
$\varepsilon\tilde{G}-\langle\varepsilon\tilde{G}\rangle$ in
(\ref{general4}) and get for the fluctuating part of the Green
function
\begin{gather}
\tilde{G}(q,q^{'})
=-F^{-1}(q)\frac{\omega^{2}_{b}k_{\perp}^{4}c_{s}^{4}}{(k_{\perp}^{2}c_{s}^{2}-\omega^{2})}
\int \varepsilon(k_{z,1})\langle
G(\omega,\mathbf{k}_{\perp},k_{z,2},q^{'})\rangle
\nonumber \\
\times\,\delta (k-k_{z,1}-k_{z,2})\,dk_{z,1}dk_{z,2}
 \label{general5}
\end{gather}
Inserting this expression into equation (\ref{general3}) yields
\begin{gather}
F(q)\langle G(q,q^{'})\rangle
-\frac{\omega^{4}_{b}k_{\perp}^{8}c_{s}^{8}}{(k_{\perp}^{2}c_{s}^{2}-\omega^{2})^{2}}\int
\langle\varepsilon(k_{z,1})\varepsilon(k_{z,3})\rangle
F^{-1}(\omega,\mathbf{k}_{\perp},k_{z,2}) \nonumber
\\
\times\langle G(\omega,\mathbf{k}_{\perp},k_{z,4},q^{'})\rangle
\delta (k-k_{z,1}-k_{z,2}) \delta (k_{z,2}-k_{z,3}-k_{z,4})
\nonumber
\\
\times \, dk_{z,1}dk_{z,2}dk_{z,3}dk_{z,4}=\delta (q-q^{'}).
\label{general6}
\end{gather}
Due to homogeneity of the random process, the correlator
(\ref{correl}) in the wave number domain has the form
\begin{equation}
\label{W-k}
\langle\varepsilon(k_{z})\varepsilon(k_{z}^{'})\rangle=W(k_{z})\delta
(k_{z}+k_{z}^{'}),
\end{equation}
and the Green function in (\ref{general6}) has the structure
$G(q,q^{'})=G(q)\delta (q-q^{'})$. Making use of this, one can
perform some integrations in (\ref{general6}), and for the mean
Green function $\langle G(q)\rangle$ we finally obtain
\begin{equation}
\label{Green-final} \langle
G(\omega,\mathbf{k}_{\perp},k_{z})\rangle=
\left[F(\omega,\mathbf{k}_{\perp},k_{z})
+\frac{\omega^{4}_{b}k_{\perp}^{8}c_{s}^{8}}{(k_{\perp}^{2}c_{s}^{2}-\omega^{2})^{2}}
\int\frac{W(k_{z}-k_{z}^{'})}{F(\omega,\mathbf{k}_{\perp},k_{z}^{'})}dk_{z}^{'}\right]^{-1}
\end{equation}
The poles of the mean Green function $\langle
G(\omega,\mathbf{k}_{\perp},k_{z})\rangle$ determine the spectrum
of elementary excitations and the corresponding dispersion
relation is
\begin{equation}
\label{disp-general} F(\omega,\mathbf{k}_{\perp},k_{z})
+\frac{\omega^{4}_{b}k_{\perp}^{8}c_{s}^{8}}{(k_{\perp}^{2}c_{s}^{2}-\omega^{2})^{2}}
\int\frac{W(k_{z}-k_{z}^{'})}{F(\omega,\mathbf{k}_{\perp},k_{z}^{'})}dk_{z}^{'}=0.
\end{equation}
In the absence of random inhomogeneity, we recover the  dispersion
relation (\ref{disp-free}). In many physical applications, the
noise spectrum $W(k)$ is well described by the Lorentz spectrum
\begin{equation}
\label{corr-Lorenz} W(k)=\frac{D}{\pi l_{c}
[1/l_{c}^{2}+(k-k_{0})^{2}]},
\end{equation}
or the Gaussian spectrum
\begin{equation}
\label{corr-Gauss}
W(k)=\frac{Dl_{c}}{2\sqrt{\pi}}\mathrm{e}^{-(k-k_{0})^{2}l_{c}^{2}/4},
\end{equation}
where $l_{c}$ is the correlation length, and $k_{0}$ is the
characteristic noise scale, but the calculation of the integral in
(\ref{disp-general}) with spectrum $W(k)$ of the form
(\ref{corr-Lorenz}) or (\ref{corr-Gauss}) is outside the scope of
this paper. In what follows, we will consider two simplest cases.
The first case corresponds to  $\delta$- correlated in $z$
autocorrelation function (\ref{correl})
\begin{equation}
\label{corr-white} B(z-z^{'})=B_{0}\delta (z-z^{'}),
\end{equation}
where $\delta (z)$ is the Dirac delta function, that is white
noise and zero correlation length $l_{c}$ \cite{Kampen}. In this
case $W(k)$ does not depend on $k$,
\begin{equation}
\label{corr-white1} W(k)=Dc_{s}/\omega_{b},
\end{equation}
so that all harmonics with the same intensity are present in the
noise spectrum. Note that $W(k)$, as can be seen from (\ref{W-k}),
has the dimension of length, and we have introduced the
dimensionless parameter $D$ by scaling to the natural length
$c_{s}/\omega_{b}$.

In the second case (the opposite of the first one), we assume that
the spectrum of noise is concentrated near a certain
characteristic scale $k_{0}$ and has the form
\begin{equation}
\label{corr-delta} W(k)=\frac{D}{2}[\delta (k-k_{0})+\delta
(k+k_{0})],
\end{equation}
where $W(k)$ in (\ref{corr-delta}) already has the dimension of
length. Note that in the limit $1/l_{c}\rightarrow 0$, the
functions (\ref{corr-Lorenz}) and (\ref{corr-Gauss}) become delta
functions, so that expression (\ref{corr-delta}) corresponds to an
infinite correlation length.

As noted above, in the first case $W(k)$ in (\ref{disp-general})
does not depend on $k$, and the integral in (\ref{disp-general})
can  easily be calculated. The two poles of the integrand in
(\ref{disp-general}) cannot lie on the real axis of the complex
$k_{z}^{'}$-plane, since otherwise this would lead to a
contradiction. Indeed, in that case the contribution of the
principal value part of the integral leads, as can be seen, to
complex $\omega$. Then the calculation of the integral gives the
dispersion relation
\begin{equation}
\label{disp-white} F(\omega,\mathbf{k}_{\perp},k_{z}) =\frac{\pi D
\omega^{3}_{b}k_{\perp}^{8}c_{s}^{7}}{(k_{\perp}^{2}c_{s}^{2}-\omega^{2})^{2}\omega^{2}
\sqrt{(\omega^{2}_{a}/c_{s}^{2}+k^{2}_{\perp})-\omega^{2}/c^{2}_{s}-\omega^{2}_{b}
k^{2}_{\perp}/\omega^{2}}}.
\end{equation}
Equation (\ref{disp-white}) for $\omega$ can be reduced to an
algebraic equation with real coefficients, so its roots represent
a pair of conjugate values. In this case, the dispersion relation
(\ref{disp-white}) predicts  instability of acoustic-gravity
waves.

In the second case, substituting (\ref{corr-delta}) into
(\ref{disp-general}), one can obtain the dispersion relation
\begin{equation}
\label{disp-param} F(\omega,\mathbf{k}_{\perp},k_{z})
+\frac{D}{2}\frac{\omega^{4}_{b}k_{\perp}^{8}c_{s}^{8}}{(k_{\perp}^{2}c_{s}^{2}-\omega^{2})^{2}}
\left[\frac{1}{F(\omega,\mathbf{k}_{\perp},k_{z}+k_{0})}+\frac{1}
{F(\omega,\mathbf{k}_{\perp},k_{z}-k_{0})}\right]=0.
\end{equation}
Equation (\ref{disp-param}) resembles dispersion equations for
parametric instabilities in plasmas \cite{Sudan1984}. On the other
hand, equation (\ref{main}) with a multiplicative random term can,
in a certain sense, be considered as a model of a stochastic
oscillator, which, as is known, can be parametrically excited. In
the next section, we show that, under certain conditions, there is
indeed an instability of acoustic-gravity waves.

\section{Instability of the gravity and acoustic waves}

Taking into account (\ref{AGW-plus-minus}), the function
$F(\omega,\mathbf{k}_{\perp},k_{z})$ can be written in the form
\begin{equation}
\label{F1}
F(\omega,\mathbf{k}_{\perp},k_{z})=[\omega_{+}^{2}(\mathbf{k}_{\perp},k_{z})
-\omega^{2}][\omega_{-}^{2}(\mathbf{k}_{\perp},k_{z})-\omega^{2}].
\end{equation}
We consider separately the low-frequency $\omega_{-}$ and
high-frequency $\omega_{+}$ branches, assuming that for the
low-frequency waves $\omega^{2}\ll \omega_{+}^{2}$, while for the
high-frequency ones $\omega^{2}\gg \omega_{-}^{2}$, so that these
two branches are sufficiently separated from each other (in
particular, this situation always occurs if $k_{z}\gg k_{\perp}$).
Based on these assumptions, from (\ref{disp-free}) we can write
for the gravity (low-frequency) waves
\begin{equation}
\label{omega-l} \omega_{-}^{2}(\mathbf{k}_{\perp},k_{z})\sim
\omega_{L}^{2}(\mathbf{k}_{\perp},k_{z})=\frac{\omega_{b}^{2}k_{\perp}^{2}c_{s}^{2}}
{k^{2}c_{s}^{2}+\omega_{a}^{2}},
\end{equation}
and for the acoustic (high-frequency) waves
\begin{equation}
\label{omega-h} \omega_{+}^{2}(\mathbf{k}_{\perp},k_{z})\sim
\omega_{H}^{2}(\mathbf{k}_{\perp},k_{z})=k^{2}c_{s}^{2}+\omega_{a}^{2}.
\end{equation}
In the following, we introduce the notations
\begin{gather}
\omega_{L}^{2}=\omega_{L}^{2}(\mathbf{k}_{\perp},k_{z}), \quad
\quad \omega_{H}^{2}=\omega_{H}^{2}(\mathbf{k}_{\perp},k_{z}),
\\
\omega_{L,\pm}^{2}=\omega_{L}^{2}(\mathbf{k}_{\perp},k_{z}\pm
k_{0}) , \quad \quad
\omega_{H,\pm}^{2}=\omega_{H}^{2}(\mathbf{k}_{\perp},k_{z}\pm
k_{0}).
\end{gather}

\subsection{White noise case}

First of all, we note that, as can be seen from
(\ref{disp-white}), the expression under the radical in
(\ref{disp-white}) is always negative near the eigenfrequencies
$\omega_{\pm}$ if $k_{z}\neq 0$, and then (\ref{disp-white})
predicts an instability of the gravity and acoustic waves.

For the gravity waves we assume
$\omega^{2}\ll\omega_{+}^{2}(\mathbf{k}_{\perp},k_{z})$ and,
taking into account (\ref{F1}), the dispersion relation
(\ref{disp-white}) becomes
\begin{equation}
\label{disp-white1} \omega^{2}_{H} (\omega^{2}_{L}-\omega^{2})
=\frac{\pi D
\omega^{3}_{b}k_{\perp}^{8}c_{s}^{8}}{(k_{\perp}^{2}c_{s}^{2}-\omega^{2})^{2}\omega^{2}
\sqrt{\omega_{H}^{2}-\omega^{2}_{H}\omega^{2}_{L}/\omega^{2}-k_{z}^{2}c_{s}^{2}}}.
\end{equation}
Then we set $\omega^{2}=\omega_{L}^{2}+\delta$ in
(\ref{disp-white1}), where $|\delta|\ll\omega_{L}^{2}$. If
$k^{2}_{z}c^{2}_{s}\gg (\omega^{2}_{H}/\omega^{2}_{L})|\delta|$,
taking into account (\ref{omega-l}) and (\ref{omega-h}), after
direct calculation one can obtain the instability growth rate
$\gamma=|\mathrm{Im}\,\delta|/(2\omega_{L})$ of gravity waves in
the form
\begin{equation}
\label{white1} \gamma=\frac{\pi
D\omega_{b}^{3}\omega_{H}^{2}k_{\perp}^{4}c_{s}^{4}}{2\omega_{L}^{3}k_{z}c_{s}
[k^{2}c_{s}^{2}+\omega_{a}^{2}-\omega_{b}^{2}]^{2}}.
\end{equation}
In the case $k_{z}=0$ (a more exact condition has the form
$k^{2}_{z}c^{2}_{s}\ll (\omega^{2}_{H}/\omega^{2}_{L})|\delta|$ )
one can write
\begin{equation}
\label{white2} \omega_{H}^{4}\delta^{3}=\frac{(\pi D
\omega_{b}^{3}k_{\perp}^{8}c_{s}^{8})^{2}}{[(k_{\perp}^{2}c_{s}^{2}-\omega_{L}^{2})^{2}
\omega_{H}\omega_{L}]^{2}},
\end{equation}
and then we have for the instability growth rate of gravity waves
\begin{equation}
\label{white3} \gamma=\frac{\sqrt{3}}{4\omega_{L}}\left[\frac{\pi
D\omega_{H}\omega_{b}^{3} k^{4}_{\perp}c_{s}^{4}
}{\omega_{L}(k^{2}c_{s}^{2}+\omega_{a}^{2}-\omega_{b}^{2})^{2}}\right]^{2/3}.
\end{equation}

Similarly, for the acoustic waves, assuming
$\omega^{2}\gg\omega_{-}^{2}(\mathbf{k}_{\perp},k_{z})$, from
(\ref{disp-white}) one  gets
\begin{equation}
\label{white-1acous}\omega
^{2}(\omega^{2}-\omega^{2}_{H})=\frac{\pi D
\omega_{b}^{3}k_{\perp}^{8}c_{s}^{8}}{(k_{\perp}^{2}c_{s}^{2}-\omega^{2})^{2}\omega^{2}
\sqrt{\omega_{H}^{2}-\omega^{2}-k_{z}^{2}c_{s}^{2}}},
\end{equation}
and setting $\omega^{2}=\omega_{H}^{2}+\delta$ in
(\ref{disp-white}), where $|\delta|\ll\omega_{H}^{2}$, one can
find the instability growth rate of acoustic waves
\begin{equation}
\label{white-2acous} \gamma=\frac{\pi
D\omega_{b}^{3}k_{\perp}^{8}c_{s}^{8}}{2\omega_{H}^{5}k_{z}c_{s}(k_{z}^{2}c_{s}^{2}+\omega_{a}^{2})^{2}}
\end{equation}
provided that $|\delta|\ll k^{2}_{z}c^{2}_{s}$. In the opposite
case $|\delta|\gg k^{2}_{z}c^{2}_{s}$, we have
\begin{equation}
\label{white-3acous}
\gamma=\frac{\sqrt{3}}{4\omega_{H}}\left[\frac{\pi
D\omega_{b}^{3}k_{\perp}^{8}c_{s}^{8}}{(k_{z}^{2}c_{s}^{2}+\omega_{a}^{2})^{2}
\omega_{H}^{4}}\right]^{2/3}.
\end{equation}
Comparison of (\ref{white1}) and (\ref{white-2acous}) implies, in
particular, that for sufficiently long perpendicular wave numbers
$k_{\perp}c_{s}/\omega_{b}\ll 1$ and the same values of the
mean-square amplitude of random inhomogeneities $D$, the
instability growth rate of the acoustic waves is much smaller than
the instability growth rate of the gravity waves. In both cases,
the instability has no threshold.

\subsection{Case of $\delta$-shaped spectrum}

\begin{figure}
\centering
\includegraphics[width=3.5in]{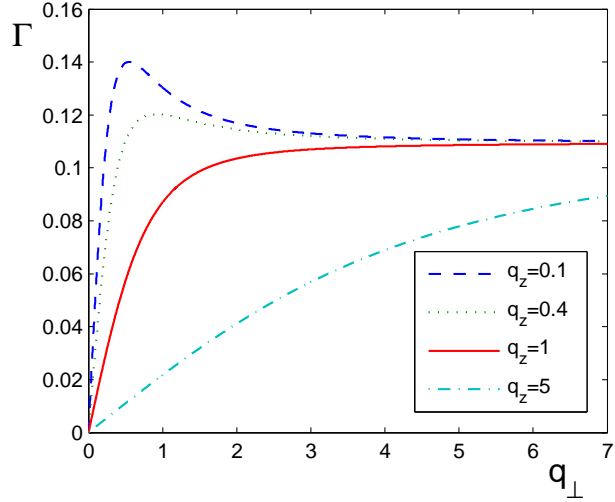}
\caption{\label{fig1} The instability growth rate $\Gamma$ of the
gravity waves v.s. the perpendicular wave number $q_{\perp}$ for
different values of the parallel wave number $q_{z}$, and $D=0.1$
in the dimensionless variables (\ref{dimensionless}).}
\end{figure}

\begin{figure}
\centering
\includegraphics[width=3.5in]{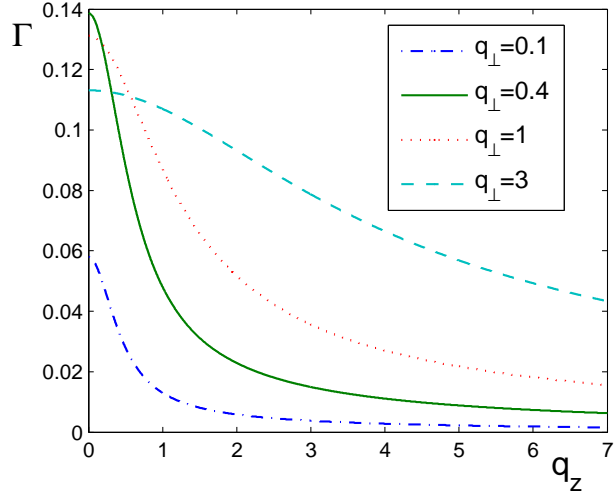}
\caption{\label{fig2} The instability growth rate $\Gamma$ of the
gravity waves v.s. the parallel wave number $q_{z}$ for different
values of the perpendicular wave number $q_{\perp}$, and $D=0.1$
in the dimensionless variables (\ref{dimensionless}).}
\end{figure}

For the gravity waves, taking into account that
$\omega_{H,\pm}^{2}\omega_{L,\pm}^{2}=\omega_{b}^{2}k_{\perp}^{2}c_{s}^{2}$,
from (\ref{disp-param}) we get the dispersion relation
\begin{equation}
\label{disp-LF}
\omega_{H}^{2}(\omega_{L}^{2}-\omega^{2})+\frac{D}{2}
\frac{\omega^{4}_{b}k_{\perp}^{8}c_{s}^{8}}{(k_{\perp}^{2}c_{s}^{2}-\omega^{2})^{2}}
\left[\frac{1}{\omega_{b}^{2}k_{\perp}^{2}c_{s}^{2}-\omega_{H,+}^{2}\omega^{2}}+\frac{1}
{\omega_{b}^{2}k_{\perp}^{2}c_{s}^{2}-\omega_{H,-}^{2}\omega^{2}}\right]=0.
\end{equation}
Setting $\omega^{2}=\omega_{L}^{2}+\delta$ in (\ref{disp-LF}),
where $|\delta|\ll\omega_{L}^{2}$, and assuming the large-scale
random inhomogeneity $k_{0}\ll k_{z}$,  we have
\begin{equation}
\label{F2} \omega_{H}^{4}\delta^{2}+
\frac{D\omega^{4}_{b}k_{\perp}^{8}c_{s}^{8}}{(k_{\perp}^{2}c_{s}^{2}-\omega^{2}_{L})^{2}}=0.
\end{equation}
Equation (\ref{F2}) predicts an instability with the growth rate
$\gamma=|\mathrm{Im}\,\omega|$ given by
\begin{equation}
\gamma=\frac{\sqrt{D}\omega_{b}k_{\perp}c_{s}\sqrt{k^{2}c_{s}^{2}
+\omega_{a}^{2}}}{2(k^{2}c_{s}^{2}
+\omega_{a}^{2}-\omega_{b}^{2})}
\end{equation}
Note that the real part of the frequency is the same as in the
absence of a random term and the frequency shift appear only in
the following order in $D$. In this case, the graphical plots of
the high- and low-frequency branches are practically similar to
the plots presented, for example, in \cite{Kaladze2008}. The
dependences of $\gamma$ on the perpendicular wave number
$k_{\perp}$ at fixed $k_{z}$ and on the parallel wave number
$k_{z}$ at fixed $k_{\perp}$  are shown in Fig.~\ref{fig1} and
Fig.~\ref{fig2}, respectively, where $D=0.1$, and we have used the
dimensionless variables
\begin{equation}
\label{dimensionless} \Gamma=\frac{\gamma}{\omega_{a}},\quad
q_{\perp}=\frac{k_{\perp}c_{s}}{\omega_{a}},\quad
q_{z}=\frac{k_{z}c_{s}}{\omega_{a}},
\end{equation}
and $\omega_{b}/\omega_{a}=1-1/\gamma\sim 0.82$ with $\gamma\sim
1.4$ for the real Earth's atmosphere. Note that, as can be seen
from Fig.~\ref{fig1}, the dependence of $\gamma$ on $k_{\perp}$ is
nonmonotonic for sufficiently small $k_{z}$. In the opposite case
of a small-scale random inhomogeneity with $k_{0}\gg k_{z}$, one
can obtain
\begin{equation}
\label{quadrat}
\omega_{H}^{2}\omega_{H,0}^{2}\delta^{2}-\omega_{H}^{2}A\delta
+\frac{D\omega^{4}_{b}k_{\perp}^{8}c_{s}^{8}}{(k_{\perp}^{2}c_{s}^{2}-\omega^{2}_{L})^{2}}=0,
\end{equation}
where
\begin{equation}
\omega_{H,0}^{2}=k^{2}_{\perp}c_{s}^{2}+k_{0}^{2}c_{s}^{2}+\omega^{2}_{a},\quad
A=\frac{\omega^{2}_{b}k_{\perp}^{2}k^{2}_{0}c_{s}^{4}}{k^{2}c_{s}^{2}+\omega^{2}_{a}}.
\end{equation}
From (\ref{quadrat}) we have
\begin{equation}
\label{quadrat1}
\delta=\frac{1}{2\omega_{H,0}^{2}}\left[A\pm\left(A^{2}
-\frac{4D\omega_{H,0}^{2}\omega^{4}_{b}k_{\perp}^{4}c_{s}^{4}}{k^{2}c_{s}^{2}
+\omega_{a}^{2}-\omega_{b}^{2}}\right)^{1/2}\right].
\end{equation}
As can be seen from (\ref{quadrat1}), an instability with the
growth rate $\gamma=|\mathrm{Im}\,\delta|/(2\omega_{L})$ appears
when
\begin{equation}
\label{threshold_grav} \frac{k^{4}_{0}c_{s}^{4}(k^{2}c_{s}^{2}
+\omega_{a}^{2}-\omega_{b}^{2})}{4(k^{2}c_{s}^{2}
+\omega_{a}^{2})^{2}(k^{2}_{0}c_{s}^{2}+k^{2}_{\perp}c_{s}^{2}
+\omega_{a}^{2})}<D.
\end{equation}
Since it is assumed that $D\ll 1$, the instability condition
(\ref{threshold_grav}) is rather strict and certainly cannot be
satisfied, for example, for $k_{0}^{2}c_{s}^{2}\gg\omega_{a}^{2}$,
since then the condition (\ref{threshold_grav}) takes the form
$(k_{0}^{2})/k_{z}^{2}\lesssim D$. On the other hand, in the
opposite case $k_{0}^{2}c_{s}^{2}\ll\omega_{a}^{2}$, the condition
(\ref{threshold_grav}) has the form
$k_{0}^{4}k_{z}^{2}c_{s}^{6}/(4\omega_{a}^{6})<D$ and may well be
satisfied.

For the acoustic waves, assuming
$\omega^{2}\gg\omega_{-}^{2}(\mathbf{k}_{\perp},k_{z})$, we have
$F(\omega,\mathbf{k}_{\perp},k_{z})=\omega^{2}(\omega^{2}-\omega_{H}^{2})
$ and from (\ref{disp-param}) one can obtain
\begin{equation}
\label{disp-HF} \omega^{2}(\omega^{2}-\omega_{H}^{2})+\frac{D}{2}
\frac{\omega^{4}_{b}k_{\perp}^{8}c_{s}^{8}}{(k_{\perp}^{2}c_{s}^{2}-\omega^{2})^{2}}
\left[\frac{1}{\omega^{2}(\omega^{2}-\omega^{2}_{H,+})}+\frac{1}{\omega^{2}
(\omega^{2}-\omega^{2}_{H,-})}\right]=0.
\end{equation}
\begin{figure}
\centering
\includegraphics[width=3.5in]{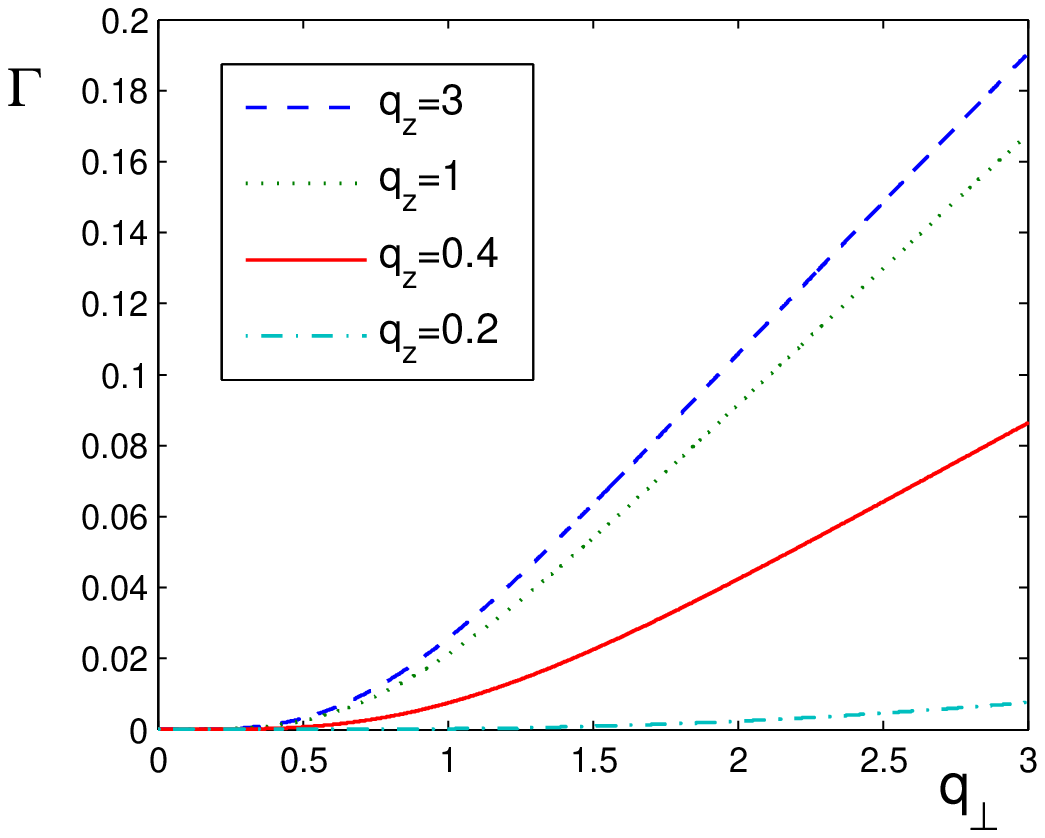}
\caption{\label{fig3} The instability growth rate $\Gamma$ of the
acoustic waves v.s. the parallel wave number $q_{z}$ for different
values of the perpendicular wave number $q_{\perp}$, and $D=0.1$
in the dimensionless variables (\ref{dimensionless}).}
\end{figure}
By analogy with the gravity waves, we set
$\omega^{2}=\omega_{H}^{2}+\delta$, where
$|\delta|\ll\omega_{H}^{2}$. For $k_{0}\ll k_{z}$ the dispersion
relation (\ref{disp-HF}) reduces to
\begin{equation}
\label{eq43}
\omega_{H}^{2}\delta+\frac{D\omega^{4}_{b}k_{\perp}^{8}c_{s}^{8}}
{\omega_{H}^{2}\delta(k_{\perp}^{2}c_{s}^{2}-\omega^{2}_{H})^{2}}=0,
\end{equation}
and describes an instability  with the growth rate
\begin{equation}
\gamma=\frac{\sqrt{D}\omega_{b}^{2}k_{\perp}^{4}c_{s}^{4}}
{(k^{2}c^{2}_{s}+\omega_{a}^{2})^{3/2}(k^{2}_{z}c^{2}_{s}+\omega_{a}^{2})}.
\end{equation}
Dependence of $\gamma$ on the parallel wave number $k_{z}$ for
different values of the perpendicular wave number $k_{\perp}$ is
shown in Fig.~\ref{fig3}, where the dimensionless variables
(\ref{dimensionless}) are used. In the case of a small-scale
inhomogeneity $k_{0}\gg k_{z}$ we have from (\ref{disp-HF})
\begin{equation}
\omega_{H}^{4}\delta^{2}-\omega_{H}^{4}k_{0}^{2}c_{s}^{2}\delta+\frac{D\omega^{4}_{b}k_{\perp}^{8}c_{s}^{8}}
{(k_{\perp}^{2}c_{s}^{2}-\omega^{2}_{H})^{2}}=0,
\end{equation}
and thus one can get
\begin{equation}
\delta=\frac{1}{2}\left\{k_{0}^{2}c_{s}^{2}
\pm\frac{1}{\omega_{H}^{2}}\left[\omega_{H}^{4}k_{0}^{4}c_{s}^{4}-\frac{4D\omega_{b}^{4}k_{\perp}^{8}c_{s}^{8}}
{(k_{z}^{2}c_{s}^{2}+\omega_{a}^{2})^{2}}\right]^{1/2}\right\}.
\end{equation}
The instability of gravity waves in this limit arises when the
threshold is exceeded
\begin{equation}
\frac{k_{0}^{4}}{k_{\perp}^{4}}<\frac{4D\omega_{b}^{4}k_{\perp}^{4}c_{s}^{4}}
{\omega_{H}^{4}(k_{z}^{2}c_{s}^{2}+\omega_{a}^{2})^{2}},
\end{equation}
and the corresponding growth rate is
\begin{equation}
\gamma=\frac{1}{4\omega_{H}^{3}}\left[\frac{4D\omega_{b}^{4}k_{\perp}^{8}c_{s}^{8}}
{(k_{z}^{2}c_{s}^{2}+\omega_{a}^{2})^{2}}-\omega_{H}^{4}k_{0}^{4}c_{s}^{4}\right]^{1/2}.
\end{equation}
Thus, in the case of a delta-shaped spectrum and small-scale
random inhomogeneities, the excitation of both gravity and
acoustic waves has a threshold character in terms of the
mean-square amplitude of random inhomogeneities.

\section{Conclusion}

Available observational data indicates a variety of possible
scenarios for the generation of AGWs in the Earth's atmosphere at
satellite measurement heights. This implies the need for a
detailed consideration of the conditions for the propagation of
these waves at such heights. In most studies of wave disturbances
in the Earth's atmosphere, the isothermal atmosphere model is
usually used, in which the equilibrium temperature is assumed to
be constant and does not change with altitude. This model is quite
realistic for sufficiently high altitudes, but even it does not
fully take into account the experimentally observed features of
atmospheric wave propagation. We have analyzed the propagation of
AGWs in the quasi-isothermal atmosphere at altitudes exceeding
$\sim 200$ km with a weakly inhomogeneous vertical temperature
profile that arises due to random non-deterministic small
perturbations. To describe small-amplitude wave disturbances in
such a region, we have derived a linear differential equation that
takes into account the weak inhomogeneity of the vertical
coordinate. For a random inhomogeneity, the resulting equation is
stochastic, and for the statistical closure procedure we have used
the Bourret approximation. The pole of the mean Green's function
of the corresponding averaged equation determines the dispersion
relation of the AGWs taking into account the weak random
inhomogeneity of the vertical temperature profile. The resulting
dispersion equation is greatly simplified in two particular cases.
The first case corresponds to random inhomogeneities in the form
of white noise, that is, a delta-correlated process in space, and
the second corresponds to a narrow spectrum of random
inhomogeneities concentrated near a certain wave number. It turned
out that both acoustic and gravity waves can be unstable, and the
corresponding growth rates were found. Note that in the isothermal
atmosphere model unstable AGWs are excluded from consideration,
since there are no sources of instabilities. In the framework of
the quasi-isothermal model we use, unstable AGWs are possible,
since the model takes into account the influence of external
random small perturbations on the equilibrium parameters of the
medium. The physical mechanism of instability is the parametric
instability of three resonantly interacting waves (see, for
example a review \cite{Staquet2002}), two of which are
high-frequency (acoustic branch), and one is low-frequency
(gravity branch).

Note that the proposed approach is quite general and includes, in
particular, the case when the vertical temperature profile
randomly depends on time as well, which naturally takes into
account the presence of turbulent temperature fluctuations. In
this case, the autocorrelation function of the random additive has
the form,
\begin{equation}
\label{correl1}
\langle\varepsilon(z,t)\varepsilon(z^{'},t^{'})\rangle=B(z-z^{'})C(t-t^{'}),
\end{equation}
and in the dispersion relation (\ref{disp-general}), terms
including $\omega-\omega^{'}$ appear in the integrand, where the
integration is carried out over $\omega^{'}$. Accordingly, in the
equation (\ref{disp-param}),  terms containing
$\omega\pm\omega_{0}$ appear. This case, as well as the analysis
of equation (\ref{disp-general}) with realistic autocorrelation
functions (\ref{corr-Lorenz}) and (\ref{corr-Gauss}), will be
considered in future works.

\section{Acknowledgments}

The work was supported by the National Research Foundation of
Ukraine, grant 2020.02/0015 "Theoretical and experimental studies
of global disturbances of natural and technogenic origin in the
Earth-atmosphere-ionosphere system". O.K.C. was also supported by
the Thematic Program of the Wolfgang Pauli Institute "Models in
Plasma, Earth and Space Sciences".



\section*{References}

\begin{thebibliography}{59}

\bibitem{Eckart1960}
C.~Eckart, Hydrodynamics of Oceans and Atmospheres, Pergamon, New
York, 1960.

\bibitem{Hines1960}
C.~O. Hines, Internal atmospheric gravity waves at ionospheric
heights, Can. J.
  Phys. 38 (1960) 1441--1481.

\bibitem{Tolstoy1967}
I.~Tolstoy, Long-period gravity waves in the atmosphere, J.
Geophys. Res. 72
  (1967) 4605--4610.

\bibitem{Liu1974}
K.~C. Yeh, C.~H. Liu, Acoustic-gravity waves in the upper
atmosphere, Rev.
  Geophys. Space Phys. 12 (1974) 193--216.

\bibitem{Beer1974}
T.~Beer, Atmospheric Waves, John Wiley, New York, 1974.

\bibitem{Gill1982}
A.~E. Gill, Atmosphere-ocean dynamics, Academic Press, New York,
1982.

\bibitem{Gossard1975}
E.~E. Gossard, W.~H. Hooke, Waves in the Atmosphere: Atmospheric
Infrasound and
  Gravity Waves: Their Generation and Propagation, Elsevier Scientific
  Publishing Company, 1975.

\bibitem{Sutherland2015}
B.~R. Sutherland, Internal Gravity Waves, Cambridge University
Press,
  Cambridge, 2015.

\bibitem{Tolstoy1963}
I.~Tolstoy, The theory of waves in stratified fluids including the
effect of
  gravity and rotation, Rev. Mod. Phys. 35 (1963) 207--230.

\bibitem{Francis1975}
S.~H. Francis, Global propagation of atmospheric gravity waves: A
review, J.
  Atmos. Sol.-Terrestrial Phys. 37 (1975) 1011-1054.

\bibitem{Collins1999}
J.~F. Lingevitch, M.~D. Collins, W.~L. Siegmann, Parabolic
equations for
  gravity and acousto-gravity waves, J. Acoust. Soc. Am. 105 (1999) 3049-3056.

\bibitem{Walter2003}
R.~L. Waltercheid, J.~H. Hecht, A reexamination of evanescent
acoustic-gravity
  waves: Special properties and aeronomical significance, J. Geophys. Res. 108
  (2003) 4340-4352.

\bibitem{Cheremnykh2021}
O.~K. Cheremnykh, A.~K. Fedorenko, Y.~A. Selivanov, S.~O.
Cheremnykh,
  Continuous spectrum of evanescent acoustic-gravity waves in an isothermal
  atmosphere, MNRAS 503 (2021) 5545-5553.

\bibitem{Kaladze2008}
T.~D. Kaladze, O.~A. Pokhotelov, H.~A. Shan, M.~I. Shan,
L.~Stenflo,
  Acoustic-gravity waves in the earth ionosphere, J. Atmos. Sol.-Terrestrial
  Phys. 70 (2008) 1607-1616.

\bibitem{Stenflo2009}
L.~Stenflo, P.~K. Shukla, Nonlinear acoustic gravity wave, J.
Plasma Phys. 75
  (2009) 841-847.

\bibitem{Izvekova2015}
Y.~N. Izvekova, S.~I. Popel, B.~B. Chen, Nonlinear
acoustic-gravity waves and
  dust particle redistribution in Earth's atmosphere, J. Atmos.
  Sol.-Terrestrial Phys. 134 (2015) 41-46.

\bibitem{Misra2019}
A.~Roy, S.~Roy, A.~P. Misra, Dynamical properties of
acoustic-gravity waves in
  the atmosphere, J. Atmos. Sol.-Terrestrial Phys. 186 (2019) 78-81.

\bibitem{Gavrilov2021}
S.~P. Kshevetskii, Y.~A. Kurdyaeva, N.~M. Gavrilov, Spectra of
acoustic-gravity
  waves in the atmosphere with a quasi-isothermal upper layer, Atmosphere 12
  (2021) 818-831.

\bibitem{Cheremnykh2022}
O.~Cheremnykh, T.~Kaladze, Y.~A. Selivanov, S.~Cheremnykh,
Evanescent
  acoustic-gravity waves in a rotating stratified atmosphere, Adv. Space Res.
  69 (2022) 1272-1280.

\bibitem{Gavrilov2018}
N.~M. Gavrilov, S.~P. Kshevetskii, A.~V. Koval, Propagation of
non-stationary
  acoustic-gravity waves at thermospheric temperatures corresponding to
  different solar activity, J. Atmos. Sol.-Terrestrial Phys. 172 (2018)
  100-106.

\bibitem{Picone2002}
J.~M. Picone, A.~E. Hedin, D.~P. Drob, A.~C. Aikin, NRLMSISE-00
{E}mpirical
  model of the atmosphere: statistical comparisons and scientific isssues, J.
  Geophys. Res. 107 (2002) 1468.

\bibitem{Burre1962}
R.~C. Bourret, Stochastically perturbed fields, with applications
to wave
  propagation in random media, Nuovo Cimento 26 (1962) 1-31.

\bibitem{Kravtsov1989}
Y.~Kravtsov, S.~Rytov, V.~Tatarskii, Principles of Statistical
Radiophysics,
  Springer-Verlag, Berlin, 1989.

\bibitem{Pedlosky1986}
J.~Pedlosky, Geophysical Fluid Dynamics, Springer-Verlag, New
York, 1987.

\bibitem{Misra_Coriolis2021}
D.~Chatterjee, A.~P. Misra, Effects of Coriolis force on the
nonlinear
  interactions of acoustic-gravity waves in the atmosphere, J. Atmos.
  Sol.-Terrestrial Phys. 222 (2021) 105722.

\bibitem{Misra_Amper1-2022}
T.~Kaladze, A.~P. Misra, A.~Roy, D.~Chatterjee, Nonlinear
evolution of internal
  gravity waves in the Earth's ionosphere: Analytical and numerical
  approach, Adv. Space. Res. 69 (2022) 3374-3385.

\bibitem{Misra_Amper2022}
A.~P. Misra, A.~Roy, D.~Chatterjee, T.~Kaladze, Internal gravity
waves in the  Earth's ionosphere, IEEE Trans. Plasma Sci. 50
(2022) 2603-2608.

\bibitem{Baydoun2015}
I.~Baydoun, D.~Baresch, R.~Pierrat, A.~Derode, Scattering mean
free path in
  continuous complex media: Beyond the Helmholtz equation, Phys. Rev. E 92
  (2015) 033201.

\bibitem{Grinevich1997}
Z.~Hryniewicz, On the range of applicability of Bourret
approximation, Appl.
  Math. Modelling 21 (1997) 247-253.

\bibitem{Kampen}
N.~G. van Kampen, Stochastic Processes in Physics and Chemistry,
Elsevier,
  Amsterdam, 2007.

\bibitem{Sudan1984}
K.~Mima, K.~Nishikawa, Parametric instabilities and wave
dissipation in
  plasmas, in: A.~A. Galeev, R.~N. Sudan (Eds.), Basic plasma physics Vol. 2,
  North Holland Publisher, New York, 1984, pp. 451-517.

\bibitem{Staquet2002}
C.~Staquet, J.~Sommeria, Internal gravity waves: From
instabilities to
  turbulence, Annu. Rev. Fluid Mech. 34 (2002) 559-593.

\end{thebibliography}
\end{document}